\documentclass[aps,pra,twocolumn, superscriptaddress,showpacs]{revtex4-1}%
\usepackage{amsfonts}
\usepackage{amsmath}
\usepackage{amssymb}
\usepackage{graphicx}%
\usepackage[english]{babel}
\usepackage{hyperref}

\usepackage[usenames,dvipsnames]{color}
\makeatletter
\@ifundefined{textcolor}{}
{%
 \definecolor{BLACK}{gray}{0}
 \definecolor{WHITE}{gray}{1}
 \definecolor{RED}{rgb}{1,0,0}
 \definecolor{GREEN}{rgb}{0,1,0}
 \definecolor{BLUE}{rgb}{0,0,1}
 \definecolor{CYAN}{cmyk}{1,0,0,0}
 \definecolor{MAGENTA}{cmyk}{0,1,0,0}
 \definecolor{YELLOW}{cmyk}{0,0,1,0}
 }

\newcommand{\beq}{\begin{equation}}
\newcommand{\eeq}{\end{equation}}

\newcommand{\ket}[1]{|#1\rangle}
\newcommand{\bra}[1]{\langle #1|}

\usepackage{babel}

\begin{document}

\title{Anti-Dynamical Casimir Effect as a Resource for Work Extraction}

\author{A. V. Dodonov
}\email{adodonov@fis.unb.br}
\affiliation{
Institute of Physics and International Center for Physics, University of Brasilia, 70910-900, Brasilia, Federal District, Brazil}

\author{D. Valente
}

\affiliation{
Instituto de F\'isica, Universidade Federal de Mato Grosso, Cuiab\'a MT, Brazil}

\affiliation{
Laboratoire Pierre Aigrain (LPA), \'Ecole Normale Sup\'erieure (ENS) - Paris, Centre National de la Recherche Scientifique (CNRS), 24 rue Lhomond, 75005 Paris, France}

\author{T. Werlang
}


\affiliation{
Instituto de F\'isica, Universidade Federal de Mato Grosso, Cuiab\'a MT, Brazil}

\begin{abstract}

We consider the quantum Rabi model with external time modulation of the atomic frequency, which can be employed to create excitations from the vacuum state of the electromagnetic field as a consequence of the dynamical Casimir effect.
Excitations can also be systematically subtracted from the atom-field system by suitably adjusting the modulation frequency, in the so-called anti-dynamical Casimir effect (ADCE). We evaluate the quantum thermodynamical work and show that a realistic out-of-equilibrium finite-time protocol harnessing ADCE allows for work extraction from the system, whose amount can be much bigger then the modulation amplitude, $| W_{\mathrm{ADCE}}| \gg \hbar \epsilon_\Omega$, in contrast to the case of very slow adiabatic modulations.
We provide means to control work extraction in state-of-the-art experimental scenarios, where precise frequency adjustments or complete system isolation may be difficult to attain.

\end{abstract}
\maketitle

%
\section{Introduction}

 The dynamical Casimir effect (DCE) consists in the generation of quanta from the initial vacuum (or any other) state of some field due to time-dependent boundary conditions or varying material properties of some macroscopic system (see \cite{book,vdodonov,revdal,nori} for reviews). 
In the majority of cases this corresponds to the generation of photons due to accelerated motion of a single mirror, a cavity wall or time-dependent dielectric permittivity/conductivity of the intracavity medium \cite{e1,e2,e3,e4,e5,e6,e7}, although phonon generation is also possible in such systems as Bose-Einstein condensates \cite{e8,e8a}, quantum fluids of light \cite{e9} or laser-cooled atomic gases \cite{e10}.
In recent years, it has been shown that a microscopic analogue of DCE can be achieved using a time-dependent quantum Rabi model \cite{r1,r2,r3} -- a quantized single-mode electromagnetic field coupled to a two-level atom (TLA) with time-dependent parameters.
The photon generation occurs as a result of time-variation of the transition frequency of the TLA or the atom-field coupling strength, while the macroscopic boundary conditions for the field remain stationary \cite{jpcs,liberato,jpa}.
Ultimately, the photon creation relies on the presence of the counter-rotating terms (CRT) in the Rabi Hamiltonian (RH) \cite{special}, usually neglected under the rotating wave approximation (RWA) \cite{jc1,jc2}.
The time-dependent Rabi model can be implemented experimentally in the current circuit Quantum Electrodynamics (circuit QED) architecture \cite{cir2,majer,ge,ger,ger1,v1,v2,v3,blais-exp,simmonds} with artificial superconducting atoms coupled to microwave resonators, where DCE analogues were already observed both in a single-mirror and a cavity configurations \cite{dce1,dce2}.
However, the DCE is not the only relevant effect that arises from the combination of the CRT and the temporal modulation of the system parameters. 
It was predicted in  Refs. \cite{igor,diego,lucas} that by properly adjusting the modulation frequency one can induce a coherent photon annihilation from non-vacuum states, in what became known as the {\em anti-dynamical Casimir effect} (ADCE).
Moreover, by letting the modulation frequency to slowly change with time, effective {\em Landau-Zener} transitions may occur between the dressed states of the time-independent Rabi Hamiltonian \cite{palermo}.

The DCE and the ADCE involve, respectively, creation and annihilation of excitations from some initial state of the atom-field system.
In order to implement these physical processes an external agent is required to supply or withdraw energy from the system by means of appropriate modulation of parameters.
However, till now there has been no clear relationship between the creation (annihilation) of excitations and the amount of energy supplied (withdrawn) by the external agent.
In this scenario, the following question can be stated: {\em is it possible to use the ADCE to extract work from the atom-field system}?
In order to address this issue, one can use the framework of quantum thermodynamics \cite{Mahler04,Horodecki13,goold15,XuerebNJP} - a field of physics seeking to establish a quantum version for thermodynamic principles and processes.
Studies in quantum thermodynamics aim to introduce appropriate definitions of work and heat \cite{HanggiPRE,Mahler08,Aberg,popescu14,HanggiPRE16,deffner16,diogo}, the development of quantum thermal machines \cite{alicki79,liden10,abah,Goswami,Bergenfeldt,RoPRL14,lutzPRL15,Ro16}, the analysis of the heat transport \cite{roukes,segal11,tanase,werlangPRE14,werlangPRE15,JoulainPRL16} and the validity of the second law at the microscopic level \cite{brandao08,esposito11,lacoste12,kosloff14,BrandaoPNAS,Oppenheim15,camati16},  the study of the stochastic thermodynamics \cite{SeifertRP,cyril15} and the fluctuation theorems \cite{Jarzynski,campisi11,batalhao14,an14,brito,TalknerNatPhys}, just to name a few.
An interesting result involving the Rabi model in the context of the quantum thermodynamics was presented in Ref. \cite{benenti}.
The authors have shown that the CRT prevent the atom-field system from reaching the absolute zero temperature, even in the limit of an infinite number of cycles.

In this paper we investigate the relationship between quantum work and the creation or annihilation of excitations in the Rabi model.
Work extraction has already been investigated in the stationary regime of the Rabi model \cite{altintas}, requiring ultra-strong couplings and huge atom-field detunings.
In our case, the time-dependent Rabi model is shown to allow for finite-time work extraction even for perturbative modulation amplitudes and moderate atom-field detunings, without demanding ultra-strong couplings.
Work can be extracted either periodically or steadily, depending on whether the modulation frequency is itself constant or time-varying.

This paper is organized as follows.
The theoretical framework is presented in Sec. \ref{II}.
In Sec. \ref{IIA} the time-dependent RH is introduced, and the system dynamics under parametric modulation is elucidated. In Sec. \ref{IIB} the definitions of work and heat are delineated.
Our main result is described in Sec. \ref{III}: ADCE can be a resource for work extraction.
In Sec. \ref{IIIA} we evidence how work extraction is related to annihilation of system excitations driven by an out-of-equilibrium finite-time protocol,  implemented as a perturbative modulation of the atomic frequency.
In Sec. \ref{IIIB} we extend the method for multi-tone modulations, which are useful for realistic initial states of the system. In Sec. \ref{IIIC} we employ a time-dependent modulation frequency in order to solve two issues, namely, the need for an asymptotic finite work extraction and the challenge of finding a very fine-tuned modulation frequency. Finally, in section \ref{IIID} we investigate how the presence of dissipation may reduce the amount of extracted work in actual implementations.
Conclusions are presented in Sec. \ref{IV}, and some formal analytical derivations are summarized in the appendix \ref{append}.

\section{Model} \label{II}

\subsection{Atom-Field Interaction}\label{IIA}

The atom-field interaction is described by the time-dependent Rabi Hamiltonian (we set $\hbar=1$) \cite{jpcs,liberato,red1,red2,jpa}
\beq\label{hrabi}
H=\omega a^\dagger a +\frac{\Omega(t)}{2}\sigma_z + g(a+a^\dagger)(\sigma_+ + \sigma_-),
\eeq
where $a$ ($a^\dagger$) is the cavity annihilation (creation) operator, and $\sigma_+=\ket{e}\bra{g}$ and $\sigma_-=\sigma_+^\dagger$ are the atomic ladder operators.
Here $\ket{g}(\ket{e})$ denotes the atomic ground (excited) state and $\sigma_z=\ket{e}\bra{e}-\ket{g}\bra{g}$.
The field number states are denoted by $\ket{n}$, such that $a^\dagger a\ket{n}=n\ket{n}$.
$\omega$ is the cavity frequency, $g$ is the atom-field coupling strength and
$\Omega(t)$ is the time-dependent atomic transition frequency.
The total average number of excitations is given by $N=\mbox{Tr}\left(\rho a^\dagger a\right)+ \mbox{Tr}\left(\rho\ket{e}\bra{e}\right)$, where $\rho$ denotes the atom-field density operator.

 We assume that the atomic frequency undergoes an external multi-tone modulation as
\beq
\Omega(t) = \Omega_0 + \sum_k \epsilon_\Omega^{(k)}\sin(\eta^{(k)}(t)t+\phi^{(k)}),
\eeq
where $\eta^{(k)}(t)$ is the $k$-$th$ modulation frequency  and $\epsilon_\Omega^{(k)}$ is the $k$-$th$ modulation amplitude.
It is worth noting that this particular choice for the modulation does not restrict the generality of our results, since for the regime considered here ($g\ll\omega,\Omega$) a weak modulation of any system parameter produces similar results \cite{jpa,igor,tom}.
We suppose that the modulation frequency $\eta^{(k)}(t)$ may also slowly change as function of time.
To uncover the effects of temporal modulation on the system dynamics, we assume a perturbative regime characterized by $\epsilon_\Omega^{(k)}\ll\Omega_0$ and $\epsilon_\Omega^{(k)}\lesssim g$. Besides, for the validity of the single-mode approximation we require the inequality $|\Delta_-|\ll\omega$, where $\Delta_-=\omega-\Omega_0$ is the average field-atom detuning.
In the following we restrict our attention to the dispersive regime, $g\sqrt{n_{max}}\ll|\Delta_-|/2$, where $n_{max}$ is the maximum number of system excitations.

The time-dependent RH can be implemented in the circuit QED architecture -- area of research that investigates the interaction between the quantized Electromagnetic field confined in microwave resonators and superconducting artificial atoms composed of Josephson junctions. Originally proposed in 2004 with the target of implementing the Jaynes-Cummings model in a highly controllable environment \cite{pra-blais}, this field has expanded enormously over the past ten years, and now it embraces a plenty of experimental architectures with different kinds of multi-level artificial atoms and sophisticated assemblies of
interconnected resonators and waveguides \cite{cir2,transmon,majer,ge,ger,majer,paik,APL}. However, all the setups have in common the attribute of strong atom-field coupling and the ability to control {\it in situ} the system parameters (e.g., the atomic transition frequency or the coupling strength). The typical parameters in current circuit QED architectures read \cite{ger1,par1,par2,par3,simmonds}: $\omega/2\pi\sim5-10$\,GHz, $g/\omega\sim 10^{-2}-10^{-1}$, $|\Delta_-|/g\sim 0-20$, $\kappa/\omega\sim\gamma/\omega\sim5\times 10^{-7} - 5\times 10^{-6}$, $T_r\sim 10-50$ mK, where $\kappa$ ($\lambda$) denotes de cavity (atom) damping rate and $T_r$ is the temperature. As will be shown in the following, these values are sufficient for the realization of the protocol proposed in this paper.

For $g/\omega,\epsilon_\Omega/\Omega_0\ll 1$ the dynamics of the time-dependent Rabi model presents qualitatively different behaviors depending on the choice of the modulation frequency $\eta(t)$ \cite{jpcs,liberato,jpa,igor,diego,lucas,tom,red2}. By properly adjusting $\eta(t)$, which depends on the initial state of the system, one can resonantly couple a specific set of the dressed states (eigenstates) of the bare RH (see Appendix \ref{append} for details).
Below we resume three qualitatively different phenomena that alter the total number of excitations: the DCE, the anti-Jaynes-Cummings behavior (AJC) and  the ADCE.

The DCE regime is characterized by the creation of photon pairs from the vacuum and occurs for the modulation frequency $\eta_{\mbox{\tiny{DCE}}}\approx2\omega$. For the initial zero-excitation state $\ket{g,0}$ the number of excitations increases through the induced transitions between the states $\ket{g,0}$, $\ket{g,2}$, $\ket{g,4}$, $\ldots$, $\ket{g,2k}$, where $k$ depends on the values of $\Delta_-$, $g$ and $\epsilon_\Omega$ \cite{igor}.
The population of the atomic excited state remains approximately unchanged in this scenario. (In reality, the transitions occur between the atom-field dressed states, which in the dispersive regime can be approximated as $\ket{g,n}$ -- see appendix \ref{append}).
On the other hand, using the same initial state and $\eta_{\mbox{\tiny{AJC}}}\approx \omega+\Omega_0$, the dynamics consists of periodic transitions between the approximate states $\ket{g,0}$ and $\ket{e,1}$. So the increase in the number of quanta occurs due to the excitation of both the atom and the cavity field. Such behavior is known as AJC regime \cite{jpcs,roberto,igor,diego} or the blue-sideband transition \cite{red1,red2}.

Both regimes presented so far are responsible for the increase in the number of excitations.
However, in Refs. \cite{igor,diego,lucas,tom} the authors showed that the number of excitations can be reduced instead, in what they called ADCE.
This effect consists of coherent annihilation of two system excitations due to the approximate transition $\ket{g,n}\longleftrightarrow\ket{e,n-3}$ (for $n\ge 3$), and takes place for the modulation frequency
$\eta_{\mbox{\tiny{ADCE}}}\approx 3\omega-\Omega_0$. For constant $\eta$ the total number of system excitations presents a periodic behavior, with the typical period of oscillation on the order of $\sim 10^{-4}g^{-1}$ for realistic experimental parameters.
One way to get a reduction in the total number of excitations without a subsequent return to its initial value is to use a time-dependent modulation frequency, $\eta(t)$, swept across the expected resonance associated with the ADCE \cite{palermo}.
A steep decrease in the number of excitations is observed when $\eta(t)\approx\eta_{\mbox{\tiny{ADCE}}}$, and as $\eta(t)$ moves away from $\eta_{\mbox{\tiny{ADCE}}}$ the resonance condition is forfeit. As shown in appendix \ref{append} this process can be viewed as an effective Landau-Zener-Stueckelberg-Majorana problem \cite{landau,zener,stuck,majo}, which asserts that the asymptotic transition between the two involved states will be complete for sufficiently small $|\dot{\eta}|$.
Hence, it is possible to deterministically induce a steady decrease in the number of excitations.

\subsection{Quantum Thermodynamics}\label{IIB}

In the context of quantum thermodynamics, the internal energy of a quantum system is the average energy $U(t)=\left\langle H(t)\right\rangle=\mbox{Tr}\left(\rho(t)H(t)\right)$, where $H(t)$ is the system's Hamiltonian and $\rho(t)$ -- its density operator.
The quantum version of the first law of thermodynamics reads \cite{alicki79,goold15}
\beq\label{ienergy}
\Delta U=W+Q,
\eeq
where $W$ is the work performed by an external agent and $Q$ is the heat supplied to the system by its thermal environment.
The quantum work $W$ computed from time $t_i$ up to $t_f$ is related to the time variation of the system's Hamiltonian,
\beq\label{work}
W=\int_{t_i}^{t_f}\mbox{Tr}\left(\rho(t)\partial_tH(t)\right)dt,
\eeq
while the heat $Q$ is related to the time variation of the density operator $Q=\int_{t_i}^{t_f}\mbox{Tr}\left(\partial_t\rho(t)H(t)\right)dt$.
For an isolated quantum system there is no heat exchange between the system and its environment, $Q=0$. Therefore, the variation of the internal energy coincides with the work done on the system ($W>0$) or extracted from the system ($W<0$).

It is important to note that the work performance crucially depends on the time variation of the system's Hamiltonian.
For example, for an isolated system described by the time-independent Rabi model, Eq. (\ref{hrabi}) with $\Omega(t)=\Omega_0$, the quantum work will always be equal to zero, even with the increase in the number of excitations caused by the counter-rotating terms.
This result shows that in order to extract work from the atom-field system it is necessary that the system's Hamiltonian itself evolves over time, driven by an external agent.

\section{Results and Discussions}\label{III}

In this section we discuss the relationship between the creation or annihilation of excitations and quantum work in the Rabi model. We shall show that generation of a finite number of excitations is accompanied by positive work performed on the system, while the coherent annihilation of quanta is accompanied by negative work of the order $\sim (-2\hbar\omega)$, i.e., the energy is withdrawn by the external agent. We shall also demonstrate that energy can be added or withdrawn from the system without changing the total number of excitation (with only infinitesimal changes in $N$), however, the net amount of work is $|W|<\hbar\omega$ in this case.

\subsection{Single-tone modulations}\label{IIIA}

\begin{figure}[!htb]
\centering
\includegraphics[width=0.99\linewidth]{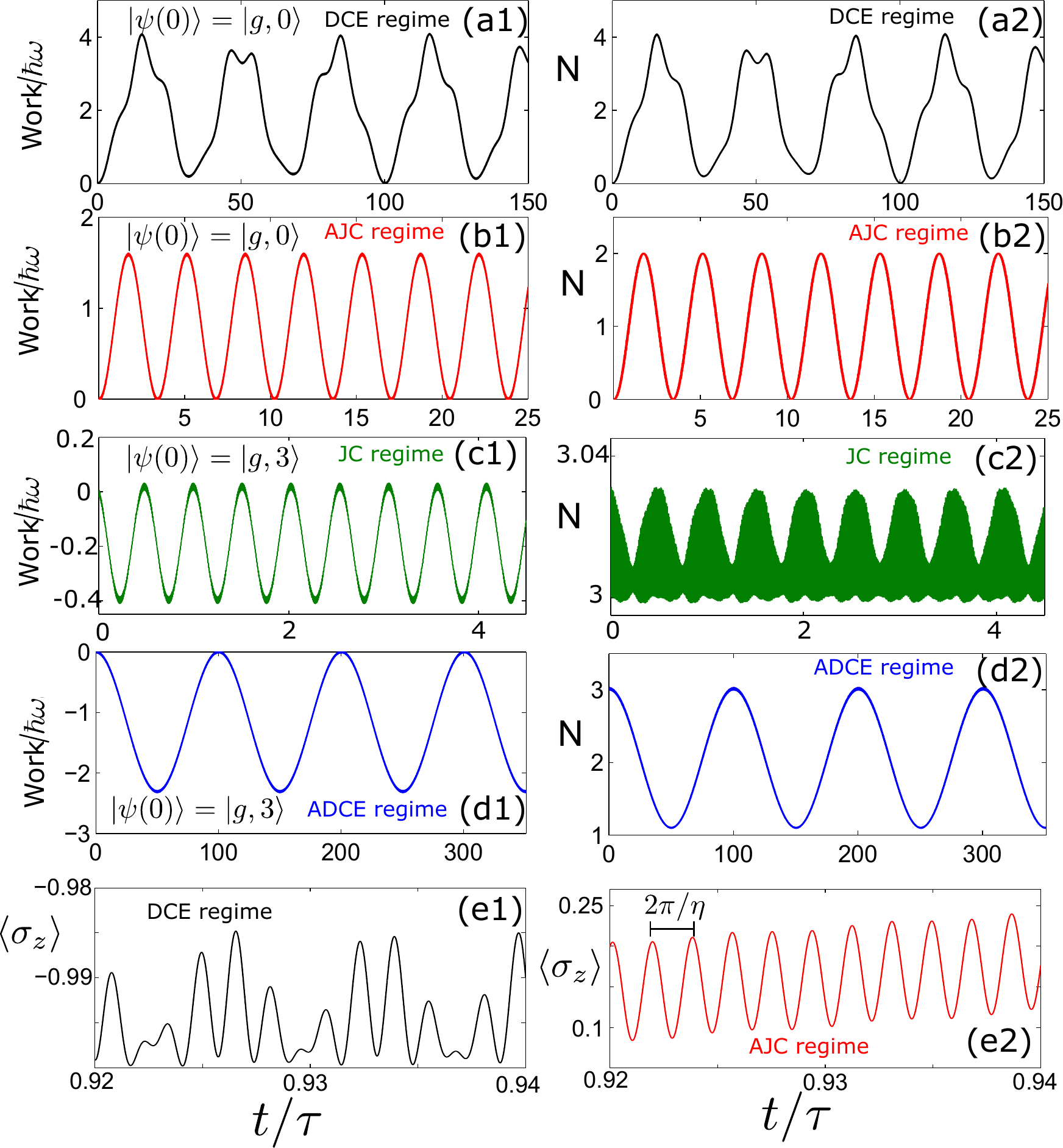}
\caption{(color online) Quantum work and average total number of excitations as a function of dimensionless time $t/\tau$ for the regimes: DCE (panels a1-a2), AJC (panels b1-b2), JC (panels c1-c2) and ADCE (panels d1-d2).
Here $\tau^{-1}=g\epsilon_\Omega/2\Delta_+$ with $\Delta_\pm=\omega\pm\Omega_0$.
For the DCE and AJC regimes the initial state is $\ket{g,0}$; for the JC and ADCE regimes it is $\ket{g,3}$.
Parameters: $g/\omega=5\times10^{-2}$, $\Delta_-=8g$, $\epsilon_\Omega=0.05\Omega_0$, $\phi=0$. The modulation frequencies are $\eta_{\mathrm{DCE}}=2.0089\omega$, $\eta_{\mathrm{AJC}}=0.9943\Delta_+$, $\eta_{\mathrm{JC}}=1.07\Delta_-$ and $\eta_{\mathrm{ADCE}}=1.0076(3\omega-\Omega_0)$. Panels e1 and e2 show the dynamics of $\langle \sigma_z\rangle$ over short timescales, exhibiting fast oscillations synchronized with $\dot{\Omega}$. In the DCE regime (panel e1) the oscillations are nonsinusoidal, but the characteristic period is still $\approx 2\pi/\eta$.
}
\label{f1}
\end{figure}

We begin our analysis by investigating how the creation of excitations from the zero-excitation state $\ket{g,0}$ affects the quantum work.
The dynamics of the system was obtained through the numerical solution of the Liouville-von Neumann equation, $\dot{\rho}=-i\left[H(t),\rho(t)\right]$, where $H(t)$ is the time-dependent RH, Eq. (\ref{hrabi}).
In Fig.\ref{f1} we plot the dynamics of the quantum work
\begin{equation}
W(t)=(1/2)\int_0^t dt' \dot{\Omega}(t')\langle \sigma_z(t') \rangle\label{megadeth}
\end{equation}
 and the total number of excitations $N$ for the parameters $g/\omega=5\times10^{-2}$, $\Delta_-=8g=0.4\omega$, $\phi=0$ and $\epsilon_\Omega=0.05\Omega_0=0.03\omega$.
The DCE regime (Fig.\ref{f1} - a1 and a2) was obtained using a single-tone modulation with $\eta=2.0089\omega$.
The AJC regime (Fig.\ref{f1} - b1 and b2) was obtained using $\eta=0.9943\Delta_+$, where $\Delta_+=\omega+\Omega_0$.
In both regimes, the quantum work is predominantly positive (it may take small negative values when $N\approx 0$) and presents an oscillatory behavior, which is periodic for AJC and quasi-periodic for DCE.
This behavior resembles the dynamics of the average number of excitations in the atom-field system.
As a positive amount of work means that an external agent does work on the system, the creation of excitations in the DCE and AJC comes from the energy supplied to the system, following the relation $W = N \hbar \omega-\hbar\Delta_-P_e$, where $P_e=\mbox{Tr}\left(\rho\ket{e}\bra{e}\right)$ is the atomic excitation probability.

Work can also be realized on the system without variation of $N$. As the simplest example, consider the Jaynes-Cummings Hamiltonian (JCH) \cite{jc1,jc2}, obtained from Eq. (\ref{hrabi}) by neglecting the counter-rotating terms. This approximate model describes well the dynamics provided $g\ll\omega,\Omega_0$ and the modulation frequency is low, $\eta\ll\omega$. For JCH the total number of excitations is a constant of motion, and for the initial state $|\varphi_{n,\pm}\rangle$, which is the eigenstate of JCH with $n$ excitations (see appendix \ref{append}), the work reads
\begin{equation}
W_{\mathrm{JC}|\eta/\omega\rightarrow0} = \pm \frac{1}{2}%
\varepsilon _{\Omega }\cos 2\theta _{n}\left[ \sin \left( \eta t+\phi
\right) -\sin \left( \phi \right) \right] .
\label{WJCsmallEta}
\end{equation}
Therefore the work of magnitude $|W_{\mathrm{JC}|\eta/\omega\rightarrow0}|\le \hbar\epsilon_\Omega$ can be performed on or by the system without any change in $N$.

Still under JCH, one can obtain a finite work, either positive or negative, by setting the modulation frequency to $\eta\approx |\Delta_-|$, thereby promoting the coupling between approximate states $|g,n\rangle$ and $|e,n-1\rangle$ in the dispersive regime. This behavior is known as Jaynes-Cummings or red-sideband regime \cite{jpcs,roberto,red1,red2}, and has recently been implemented experimentally in circuit QED \cite{blais-exp,simmonds}. For instance, for the initial state $|g,n\rangle$ a straightforward energetic 	
reasoning predicts for the extremum amount of work $W_{\mathrm{JC}} \approx-\hbar\Delta_-$, so energy can be added or withdrawn from the system depending on the sign of $\Delta_-$. This result holds under RH as well, as illustrated in Fig. \ref{f1} - c1 and c2 for the initial state $|g,3\rangle$ and the modulation frequency $\eta=1.07\Delta_-$ (all other parameters are as previously). We see that work is indeed extracted from the system, while the total number of excitations undergoes only infinitesimal changes due to the off-resonant contribution of CRT.

Because in the DCE and AJC regimes the maximal extracted work is $W_{\mathrm{max}} > |W_{\mathrm{JC}}|$, one can see that the counter-rotating terms crucially contribute to the quantum work. Since the creation of excitations is related to the performance of work on the system, the annihilation of excitations is expected to play a role in the extraction of work.
To investigate this point, we evaluate the dynamics of $W(t)$ in the ADCE regime.
For this, we adopted $\eta=1.0076(3\omega-\Omega_0)$ and the initial state $\ket{g,3}$.
Fig. \ref{f1}- d1 and d2 show that the quantum work becomes negative indeed, proving that the ADCE can be used to extract work from the atom-field system.
Maximal work extraction occurs at times when the population of the state $\ket{g,3}$ attain its minimum value due to the transfer of population to the state $\ket{e,0}$, that is, when $N$ reaches its lowest value. We also see that in the ADCE regime one can extract the energy $\approx 2\hbar\omega$ from the system, compared to $\approx \hbar\omega/2$ under the JC resonance.

The main contribution of this paper to the field of quantum thermodynamics is that it shows a finite-time, out-of-equilibrium realistic resource for work extraction from systems suitably described by the Rabi model.
The Rabi model itself has already been investigated in the context of heat engines \cite{altintas}.
In that case, the so called adiabatic regime was addressed, which restricts the rate of change in any system's parameter to an infinitely slow pace, in order to keep it in equilibrium at all points of the cycle.
As a consequence, the exchanged work is directly proportional to the variation of the system's energy gap, since populations of the eigenstates remain constant during the work exchange protocol for an isolated atom-field system.
Therefore, a fair amount of work extraction requires huge variations of the gap, that means equally huge changes in the atom-field detuning, as well as an ultra-strong coupling regime, $g\sim \omega$.
The out-of-equilibrium process we present here conveys a conceptually different origin for work extraction: it comes from the time-dependent gap variation that under resonance conditions induces the amplification of oscillations between the eigenstates of the non-modulated system, thereby allowing the system to be driven to a lower-energy state.

The quantum thermodynamics approach also leaves clear the interference nature associated with the generation or annihilation of excitations in our protocol. As is clear from Eq. (\ref{megadeth}), the work averaged over a few periods of oscillation of $\Omega$ would be zero unless $\langle \sigma _{z}(t)\rangle $ also exhibits fast oscillations with frequencies of the order of $\eta$. This is precisely what one observes by making a zoom of $\langle\sigma_z (t)\rangle$ corresponding to data in Figs. \ref{f1} - a1 and b1, as shown in Figs. \ref{f1} - e1 and e2. This confirms that a finite average amount of work can only be obtained when the oscillations of $\langle \sigma _{z}(t)\rangle $ become synchronized with $\dot{\Omega}(t)$, meaning that energy can be added or withdrawn from the system only under resonance conditions. Since the modulation frequency must also match the energy gap associated to the transition between two system eigenstates, modulation frequencies of the order $\sim 2\omega$ lead to larger amount of added or extracted work than modulation frequencies of the order $\sim|\Delta_-|$.

\subsection{ADCE under multi-tone modulation}\label{IIIB}

The initial state influences the maximum amount of work that can be extracted from the system by means of ADCE.
As a realistic example, we consider the scenario where the field is initially in a thermal state and
the atom is in the ground state, $\rho(0)=\ket{g}\bra{g}\otimes\rho^f_T$. We recall that the thermal state is given by $\rho^f_T=\sum_np_n\ket{n}\bra{n}$, where $p_n=\bar{n}^n/(\bar{n}+1)^{n+1}$ is the  population of the state $\ket{n}$ and $\bar{n}$ is the average photon number related to temperature as $T^{-1}=k_B\omega^{-1}\ln(\frac{1+\bar{n}}{\bar{n}})$ ($k_B$ is the Boltzmann constant).

\begin{figure}[!htb]
\centering
\includegraphics[width=0.99\linewidth]{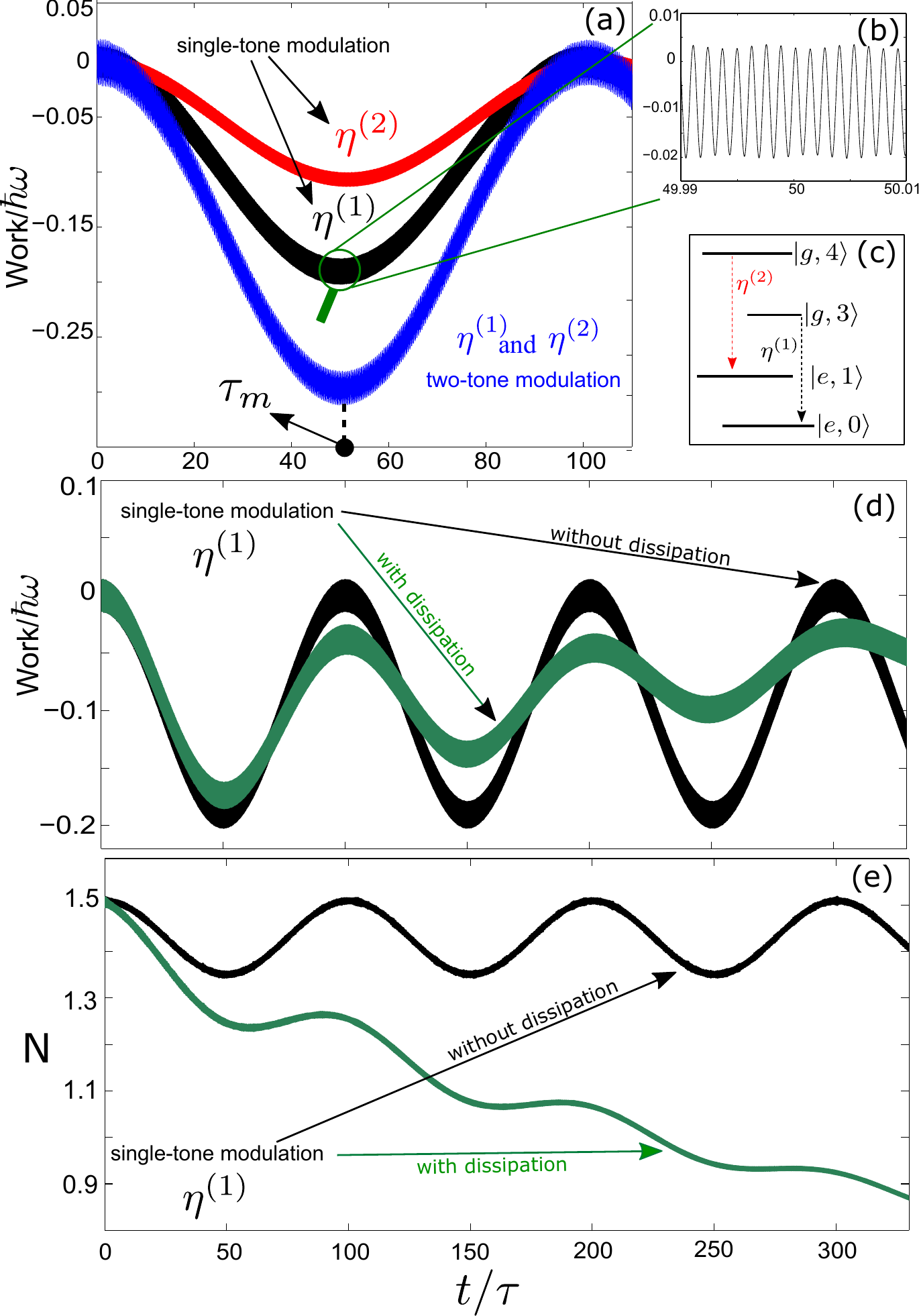}
\caption{(color online) {\it Panel a}: Quantum work as a function of dimensionless time $t/\tau$ for single-tone modulations with $\{\eta^{(1)}=1.0076(3\omega-\Omega_0),\epsilon_\Omega^{(1)}=0.05\Omega_0\}$ (black curve), $\{\eta^{(2)}=1.0113(3\omega-\Omega_0), \epsilon_\Omega^{(2)}=0.025\Omega_0\}$ (red curve), and for the two-tone modulation with the above frequencies and modulation amplitudes (blue curve). All curves correspond to the ADCE regime and $\phi^{(1,2)}=0$.
We used the initial state $\rho(0)=\ket{g}\bra{g}\otimes\rho^f_T$, where $\rho^f_T=\sum_np_n\ket{n}\bra{n}$ is a field thermal state with $p_n=\bar{n}^n/(\bar{n}+1)^{n+1}$ and $\bar{n}=1.5$ (other parameters are as in Fig. \ref{f1}).
$\tau_m$ denotes the instant of time when the extracted work is maximum.
{\it Panel b:} Dynamics of quantum work around the instant of time $\tau_m$ for a single-tone modulation with frequency $\eta^{(1)}$.
{\it Panel c:} Level diagram related to the modulation-induced transitions between the approximate eigenstates of the time-independent RH for frequencies $\eta^{(1)}$ and $\eta^{(2)}$.{\it Panel d:} Quantum work for single-tone modulation $\eta^{(1)}$ in the presence of Markovian dissipation. The cavity and atom damping rates are $\kappa=\gamma=2\times 10^{-5}g$, and the reservoirs' temperature is $k_B T_r/\omega=0.33$. {\it Panel e:} Behavior of $N$ under unitary and dissipative dynamics.}
\label{f2}
\end{figure}

\begin{figure}[!htb]
\centering
\includegraphics[width=0.99\linewidth]{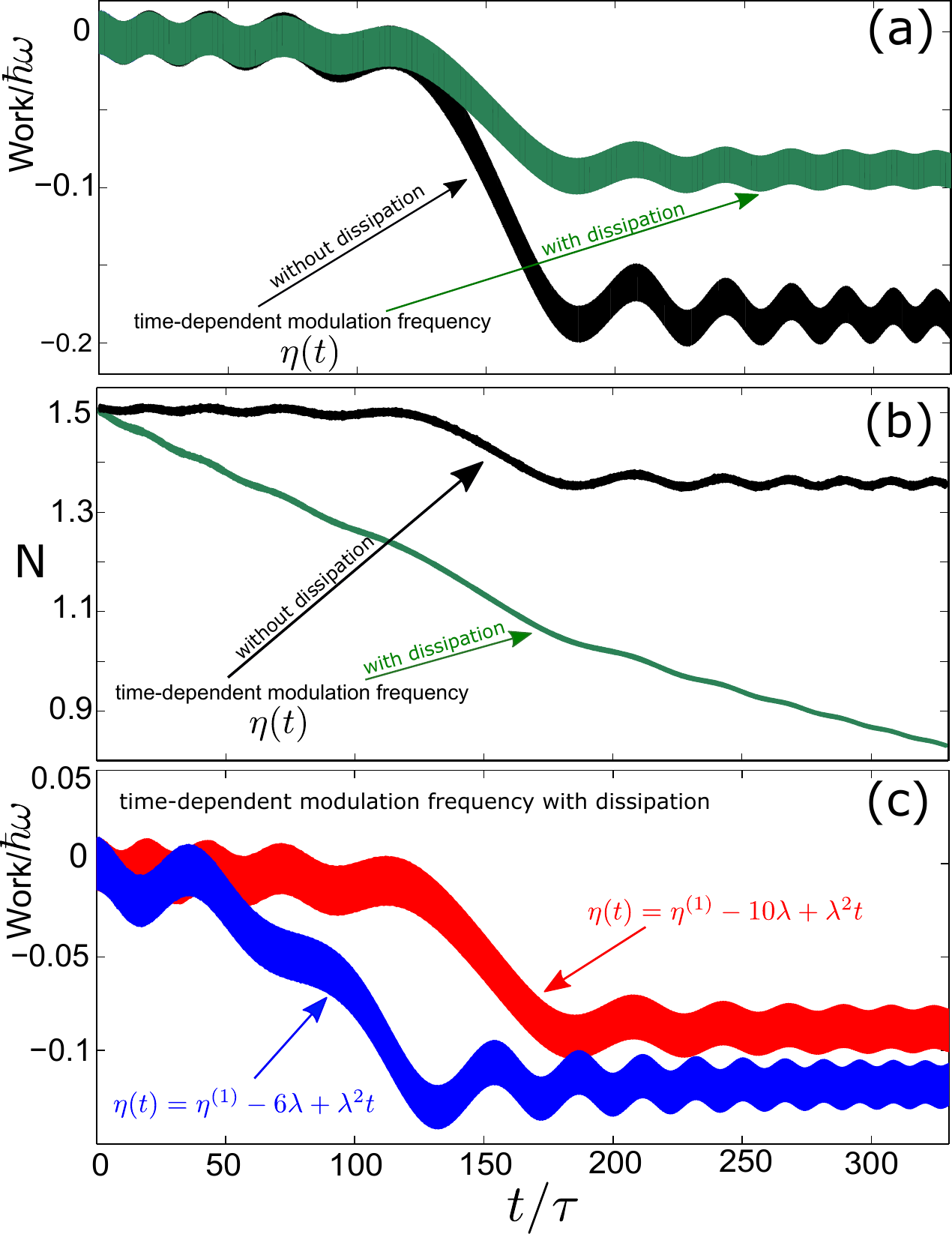}
\caption{(color online) {\it Panels a-b}: Quantum work and $N$ under unitary (black curves) and dissipative (green curves) evolution for aperiodic modulation with frequency $\eta(t)=\eta^{(1)}-10\lambda + \lambda^2t$, where
$\eta^{(1)}=1.0076(3\omega-\Omega_0)$, $\phi=0$ and $\lambda=1.67\times10^{-5}\omega$.
The initial state and other parameters are as in Fig. \ref{f2}. The dynamics resembles the typical Landau-Zener behavior for an effective TLA undergoing adiabatic frequency sweep.  {\it Panel c}: Quantum work under dissipative dynamics for two different choices of the modulation frequency: $\eta(t)=\eta^{(1)}-10\lambda + \lambda^2t$ (red curve) and $\eta(t)=\eta^{(1)}-6\lambda + \lambda^2t$ (blue curve). One can see that for the blue curve the work extraction is larger and faster, hence there is room for further optimization of our protocol.
}
\label{f3}
\end{figure}

Since the modulation frequency depends on the initial state, we can adjust the value of $\eta$ to select a particular transition $\ket{g,n}\longleftrightarrow\ket{e,n-3}$ for a given value of $n$.
Therefore, the amount of work extracted from the system depends on the initial populations of the states $\ket{g,n}$ and $\ket{e,n-3}$. As shown in the appendix \ref{append}, one is able to extract work from any initial state of the form $\ket{g}\bra{g}\otimes\rho^f$  provided the initial population $P_{\ket{g,n}}$ is larger than $P_{\ket{e,n-3}}$.
To illustrate this point we plot in Fig. \ref{f2} - a the quantum work $W$ as a function of dimensionless time $t/\tau$ for two different modulation frequencies: $\eta^{(1)}=1.0076(3\omega-\Omega_0)$ (black curve) and $\eta^{(2)}=1.0113(3\omega-\Omega_0)$ (red curve).
We use the same parameters as previously, $\bar{n}=1.5$ and $\phi^{(1,2)}=0$.
As discussed in section \ref{IIIA}, the dynamics of the quantum work in the ADCE regime presents a periodic behavior on large timescales.
In Fig. \ref{f2} - a we adopt a time interval that corresponds to a single oscillation period.
Frequency $\eta^{(1)}$ drives the transition $\ket{g,3}\longleftrightarrow\ket{e,0}$, whereas frequency $\eta^{(2)}$ drives the transition $\ket{g,4}\longleftrightarrow\ket{e,1}$.
Since the initial population of state $\ket{g,3}$ is larger than the initial population of state $\ket{g,4}$, $p_3\approx0.086>p_4\approx0.052$, it is possible to extract a larger amount of work using the modulation frequency $\eta^{(1)}$ rather than the frequency $\eta^{(2)}$~\cite{foot}.
The instant of time $\tau_m$ at which the work extraction is maximum depends on both the number of photons in the state $\ket{g,n}$ and the modulation amplitude $\epsilon_\Omega$: $\tau_m^{-1}\propto\sqrt{n(n-1)(n-2)}\epsilon_\Omega^{(k)}$ (see Appendix \ref{append} for details).
Fig. \ref{f2} - b shows the dynamics of quantum work around the instant of time $\tau_m$.
The quantum work rapidly oscillates at time scale $1/\eta^{(1)}$ due to the fast oscillations of $\langle \sigma_z\rangle$, which are necessary to withdraw a finite amount of energy from the system (as discussed in section \ref{IIIA}).

In order to adjust the same $\tau_m$ for both modulation frequencies, $\eta^{(1)}$ and $\eta^{(2)}$, the modulation amplitude associated with frequency $\eta^{(2)}$ was set as $\epsilon^{(2)}_\Omega=\epsilon^{(1)}_\Omega/2$, where $\epsilon^{(1)}_\Omega=0.05\Omega_0$ is the modulation amplitude associated with frequency $\eta^{(1)}$.
This choice of modulation amplitudes is particularly convenient for employing multi-tone modulations.
The blue curve in Fig. \ref{f2} - a describes the dynamics of the quantum work for a two-tone modulation characterized by the atomic frequency $\Omega(t) = \Omega_0 + \sum_{k=1,2} \epsilon_\Omega^{(k)}\sin(\eta^{(k)}t)$.
In this case, the amount of work extracted is effectively affected by both transitions: $\ket{g,3}\longleftrightarrow\ket{e,0}$ and $\ket{g,4}\longleftrightarrow\ket{e,1}$.
Hence, according to expression (\ref{megadeth}), the total amount of work extracted from the system is equal to the sum of the works extracted by each single-tone modulation individually.
Therefore, in order to maximize the extracted work, one needs to adjust the amplitudes $\epsilon_\Omega^{(k)}$ in such a way that all the induced transitions have the same $\tau_m$.

\subsection{Effective Landau-Zener transitions}\label{IIIC}

In the cases studied so far, we observed two features that can possibly be regarded as issues.
Firstly, the dynamics of the quantum work in the ADCE regime presents a periodic behavior.
The amount of work extracted from the system will, then, be maximal around specific instants of time ($t=n\tau_m$ with $n=1,2,\ldots$), after which it will return to zero.
Secondly, the ADCE regime is obtained for a very fine-tuned modulation frequency $\eta$, which must be ultimately found either numerically or experimentally.

We overcome these possible limitations by using a time-dependent modulation frequency $\eta(t)$ \cite{palermo}.
As discussed in section \ref{IIA}, when the modulation frequency varies over time, it can assume values close to the resonance frequency $\eta_{\mbox{\tiny{ADCE}}}$, that corresponds to the ADCE regime for a given initial state.
When $\eta(t)$ is close to $\eta_{\mbox{\tiny{ADCE}}}$, the energy of the atom-field system will decrease due to the work extraction.
But, as $\eta(t)$ moves away from $\eta_{\mbox{\tiny{ADCE}}}$, the resonance condition $\eta(t)\approx\eta_{\mbox{\tiny{ADCE}}}$ is lost.
Therefore, the external agent responsible for the atomic frequency modulation will not be able to give back energy to the system.
To illustrate this point we choose $\eta(t)=\eta^{(1)}-10\lambda + \lambda^2t$, where $\lambda$ is the transition rate between the states $\ket{g,3}$ and $\ket{e,0}$.
The initial state is the local thermal state $\rho(0)=\sum_np_n\ket{g,n}\bra{g,n}$, and all other parameters are as in section \ref{IIIB}.
In this case, $\eta^{(1)}=1.0076(3\omega-\Omega_0)$ and $\lambda=1.67\times10^{-5}\omega$.

In Fig. \ref{f3} - a  we plot the quantum work as a function of dimensionless time $t/\tau$.
The transition $|g,3\rangle \longrightarrow |e,0\rangle$ begins at $t_b\approx5/\lambda\approx 140\tau$, corresponding to $\eta(t_b)=\eta^{(1)}-5\lambda$, instead of $\eta(t_b)=\eta^{(1)}$ one would expect naively. Such discrepancy can be explained after rigorous derivation of the effective Hamiltonian, which reads $H_{\mathrm{eff}}=V(t)(\ket{g,3}\bra{g,3}-\ket{e,0}\bra{e,0})/2+(\lambda\ket{g,3}\bra{e,0}+h.c.)$, where $V(t)=2\lambda^2t-10\lambda$ (see appendix \ref{append}).
Indeed, for $t\approx t_b$ one obtains $V(t_b)\approx 0$, corresponding to the expected resonance condition. Fortunately, we do not have to solve analytically the time evolution according to the Hamiltonian $H_{\mathrm{eff}}$, since this was made independently in 1932 by \emph{Landau, Zener,  Stueckelberg and Majorana}, in what became known as Landau-Zener problem \cite{landau, zener,stuck,majo}. As shown in the appendix \ref{append}, for sufficiently slow $\dot{\eta}$ the transition from $|g,3\rangle$ to $|e,0\rangle$ is almost complete, even for finite duration of the process \cite{palermo}.
Therefore, the work extracted from the system tends to a steady value $W\approx-0.2\hbar\omega$.
Note that this steady value is practically equal to the maximum extracted work using the time-independent modulation frequency $\eta=\eta^{(1)}$, shown in Fig. \ref{f2} - a (black curve).

\subsection{Account of dissipation}\label{IIID}

For actual experimental implementation of our proposal it is necessary to take into account the interaction between the system and its environments. For the quantum Rabi model with moderate coupling rates, $g/\omega \lesssim 10^{-1}$, one can use the Markovian master equation in the dressed picture, which was rigorously deduced from the first principles in \cite{red1}. We did solve it numerically and verified that for the parameters of Figs. \ref{f1} and \ref{f2} the results are almost indistinguishable from the predictions of a much simpler \lq standard master equation\rq\ of Quantum Optics: $\dot{\rho}=-i\left[H(t),\rho(t)\right]+\mathcal{L}(\rho)$.
The Liouvillian superoperator $\mathcal{L}(\rho)$ reads \cite{pra-blais,red1,diego,palermo}
\begin{eqnarray}
\mathcal{L}(\rho)&=&\gamma(1+n_a)\mathcal{D}[\sigma_-]\rho+\gamma n_a\mathcal{D}[\sigma_+]\rho\nonumber\\
&+& \kappa(1+n_c)\mathcal{D}[a]\rho+\kappa n_c\mathcal{D}[a^\dagger]\rho,
\end{eqnarray}
where $\mathcal{D}[c]\rho\equiv\frac{1}{2}(2c\rho c^\dagger-c^\dagger c\rho-\rho c^\dagger c)$, $\kappa$ ($\gamma$) is the decay rate of the cavity (atom) and $n_c=[\exp(\beta\hbar\omega)-1]^{-1}$ ($n_a=[\exp(\beta\hbar\Omega_0)-1]^{-1}$) is the mean number of thermal excitations in the cavity (atom). Here $\beta=k_B T_r$ and $T_r$ is the common temperature of both reservoirs, adjusted so that $n_c=0.05$. We use the same parameters as previously, which give $n_a=0.19$ for the positive detuning adopted in this paper. For the dissipative rates we assume the state-of-the-art values in circuit QED: $\kappa=\gamma=2\times10^{-5}g$.

We studied the influence of dissipation on the quantum work and average number of excitations in the ADCE regime for the initial state $\rho(0)=|g\rangle\langle g|\otimes\rho_T^f$ with $\bar{n}=1.5$. In Fig. \ref{f2} - d and e we show the behavior of $W$ and $N$ for a single-tone modulation that drives the transition $|g,3\rangle \longleftrightarrow |e,0\rangle$. For initial times the amount of extracted work is approximately equal to the one obtained in the lossless case, although its absolute value decreases as time goes on. The average number of excitations decreases with exponential envelope superposed with small oscillations due to the periodic transitions between $|g,3\rangle$ and $|e,0\rangle$, which are resolvable for initial times.

The aperiodic regime with time-dependent $\eta$ is also feasible in the presence of dissipation, as illustrated in Fig. \ref{f3}. The behavior of $W$ and $N$ for the modulation frequency $\eta(t)=\eta^{(1)}-10\lambda + \lambda^2t$ is illustrated in Fig. \ref{f3} - a and b. The qualitative behavior of quantum work is similar in the unitary and dissipative cases, although the amount of extracted work is roughly $50\%$ smaller due to the losses. The behavior of $N$ is also affected by the ADCE: instead of an exponential decay expected for pure damping, $N$ exhibits an accentuated decrease around $t\sim 130\tau$, which is precisely where the LZ transition takes place.

We finally note that in the presence of dissipation it is advantageous to decrease the duration of the frequency sweep in the effective LZ process, so that the transition occurs at earlier times while the population of the state $|g,3\rangle$ is as high as possible. The downside of such a drastic measure is that the probability of complete population transfer from $|g,3\rangle$ to $|e,0\rangle$ is lowered and the Landau-Zener formula (\ref{Dau}) loses its validity. Yet a compromise can be found in order to optimize the work extraction. An example is shown in Fig. \ref{f3} - c for the modulation frequency $\eta(t)=\eta^{(1)}-6\lambda + \lambda^2t$ (blue curve): the LZ transition takes place earlier than in the previous case, and the amount of extracted works is slightly higher. This results demonstrate that our protocol can be optimized for maximum work extraction in realistic scenarios.

\section{Conclusions}\label{IV}
We have established for the first time the direct relationship between the quantum work and generation or annihilation of excitations in the quantum Rabi model with time-modulated atomic frequency. Our results are valid in the dispersive regime of light-matter interaction ($|\Delta_-|\sim 10g$), moderate coupling strengths ($g/\omega\sim 0.05$) and perturbative modulation amplitude ($\epsilon_\Omega/\Omega_0 \sim 0.05$), and can be easily extended for the modulation of other parameters. We showed that the \lq rotating\rq\ (or \lq Jaynes-Cummings\rq ) terms in the Rabi Hamiltonian can be used both do add and extract energy from the system while maintaining the total number of excitations approximately constant.
However, the maximum value of the work is limited by $|W_{\mathrm{max}}|\lesssim\hbar|\Delta_{-} |<\hbar\omega$ in this case.
Even in this number-conserving scenario, the out-of-equilibrium protocol we propose outperforms the very slow adiabatic modulations, i.e., $|W_{\mathrm{max}}| \sim 0.5 \hbar\omega  > \hbar \epsilon_\Omega = 0.03 \hbar\omega$.

Much greater amount of energy can be transferred between the system and the external agent by harnessing the \lq counter-rotating\rq\ (or \lq anti-Jaynes-Cummings\rq ) terms. We showed that the generation of excitations is accompanied by positive work, as in the dynamical Casimir and anti-Jaynes-Cummings effects. On the other hand, the work becomes negative in the so called anti-dynamical Casimir effect (ADCE), when one pair of excitations is coherently annihilated under appropriate modulation frequency. For ADCE the maximum amount of extracted work is $|W_{\mathrm{ADCE}}| \sim 2\hbar\omega \gg \hbar \epsilon_\Omega = 0.03\hbar\omega$, and strongly depends on the initial state. We also explained why the generation and annihilation of excitations are always accompanied by fast low-amplitude oscillations of the atomic population inversion: the attainment of a finite amount of work requires the synchronization of oscillations of $\langle \sigma_z(t)\rangle$ and the atomic transition frequency $\Omega(t)$.

We extended our results to realistic initial states in circuit QED, arguing that multi-tone modulations are more effective in extracting energy from the system when the field is prepared in a thermal equilibrium state. Additionally, we have shown how effective Landau-Zener transitions between the system dressed-states may be employed for obtaining aperiodic work extraction, which also solves the problem of an extremely fine-tuned adjustment of the modulation frequency. Lastly, we carried out numerical simulations to assess the feasibility of our proposal in realistic circuit QED setup subject to atomic and cavity dampings, demonstrating that periodic and aperiodic work extraction is still possible, although in a smaller amount compared to the lossless case. We hope these results will find applications in out-of-equilibrium quantum thermal machines of finite-time cycles.

\appendix
\section{Analytical approach}\label{append}

In this appendix we briefly describe how to obtain the resonant modulation
frequencies and the corresponding transition rates mentioned in the main
text. We expand the state corresponding to the Hamiltonian (\ref{hrabi}) as%
\begin{equation*}
|\psi (t)\rangle =\sum_{l}A_{l}(t)e^{-itE_{l}}|R_{l}\rangle ~,
\end{equation*}%
where $E_{l}$ and $|R_{l}\rangle $ are the eigenvalues and eigenstates (%
{\it dressed states}) of the time-independent Rabi Hamiltonian $H_{0}\equiv
H[\Omega (t)=\Omega _{0}]$, where $l$ increases with energy. The probability
amplitudes obey the differential equations%
\begin{eqnarray}
i\dot{A}_{j}&=&\sum_{l}\frac{\sum_{k}\varepsilon _{\Omega }^{(k)}\sin \left(
\eta ^{(k)}(t)t+\phi ^{(k)}\right) }{2}\langle R_{j}|\sigma
_{z}|R_{l}\rangle \nonumber \\
&&\times e^{-it(E_{l}-E_{j})}A_{l}~.  \label{dave}
\end{eqnarray}%
Therefore modulation frequency $\eta ^{(k)}=|E_{l}-E_{j}|$ may induce a
resonant coupling between the amplitudes $A_{j}$ and $A_{l}$ with transition
rate $\propto \left\vert \langle R_{j}|\sigma _{z}|R_{l}\rangle \right\vert $%
.

To diagonalize $H_{0}$ we perform the unitary transformation $U=\exp \left[
\Lambda (a\sigma _{-}-a^{\dagger }\sigma _{+})+\xi (a^{2}-a^{\dagger
2})\sigma _{z}\right] $, where $\Lambda =g_{0}/\Delta _{+}$, $\xi
=g_{0}\Lambda /2\omega $, $\Delta _{\pm }=\omega \pm \Omega _{0}$. To the
first order $\Lambda $ we obtain%
\begin{equation*}
U^{\dagger }H_{0}U= (\omega +\delta _{+}\sigma _{z})n+\frac{\Omega
_{0}+\delta _{+}}{2}\sigma _{z}+g(a\sigma _{+}+a^{\dagger }\sigma _{-})~,
\end{equation*}%
where $\delta _{\pm }=g^{2}/\Delta _{\pm }$. Hence we find the approximate
eigenvalues and eigenstates: $E_{0}= -(\Omega _{0}+\delta _{+})/2$, $%
|R_{0}\rangle = U|g,0\rangle $,%
\begin{equation*}
E_{m,\pm }=  \omega m-\frac{\omega +\delta _{+}}{2}\pm \frac{1}{2}%
\sqrt{\left( \Delta _{-}-2\delta _{+}m\right) ^{2}+4g^{2}m}
\end{equation*}%
\begin{equation*}
|R_{m,-}\rangle = U\left( \cos \theta _{m}|g,m\rangle -\sin \theta
_{m}|e,m-1\rangle \right)
\end{equation*}%
\begin{equation*}
|R_{m,+}\rangle = U\left( \sin \theta _{m}|g,m\rangle +\cos \theta
_{m}|e,m-1\rangle \right) ~,
\end{equation*}%
where $m\geq 1$ and%
\begin{equation*}
\tan \theta _{m}=\frac{\Delta _{-}-2\delta _{+}m+\sqrt{\left( \Delta
_{-}-2\delta _{+}m\right) ^{2}+4g^{2}m}}{2g\sqrt{m}}~.
\end{equation*}
The Jaynes-Cummings eigenvalues and eigenstates (denoted as $|\varphi_{n,\pm}\rangle$) are obtained simply by setting $U=1$ and $\delta_+=0$ in the above formulae.
To compute Eq. (\ref{WJCsmallEta}), we must note that $\langle \sigma_z (t) \rangle = \pm (\cos^2 \theta_m - \sin^2 \theta_m) = \pm \cos 2\theta_m$ for the initial state $\ket{\varphi_{n,\pm}}$.

In the following we restrict our attention to the dispersive regime, $%
|\Delta _{-}|/2\gg g\sqrt{n_{\max }}$, where $n_{\max }$ is the maximum
number of system excitations and we assume the condition $|\Delta _{-}|\ll \omega $%
. The approximate eigenenergies then read%
\begin{equation*}
E_{m,\mathcal{D}}\approx \left( \omega +\delta _{-}-\delta _{+}\right)
m-\alpha m^{2}+E_{0}
\end{equation*}%
\begin{equation*}
E_{m,-\mathcal{D}}\approx \left( \omega -\delta _{-}+\delta _{+}\right)
m+\alpha m^{2}-\Delta _{-}+E_{0},
\end{equation*}%
where $\mathcal{D}\equiv \Delta _{-}/|\Delta _{-}|=\pm 1$ and $\alpha
=g^{4}/\Delta _{-}^{3}$. To the first order in $g/\Delta _{-}$ the dressed
states are $|R_{0}\rangle \approx |g,0\rangle $,%
\begin{equation}
|R_{m,\mathcal{D}}\rangle \approx |g,m\rangle +\frac{g\sqrt{m}}{\Delta _{-}}%
|e,m-1\rangle  \notag
\end{equation}%
\begin{equation}
|R_{m,\mathcal{-D}}\rangle \approx |e,m-1\rangle -\frac{g\sqrt{m}}{\Delta
_{-}}|g,m\rangle ~.
\end{equation}

\emph{AJC regime}. For the initial state $|g,0\rangle $ and the modulation
frequency $\eta ^{(k)}=\Delta _{+}-2\left( \delta _{-}-\delta _{+}\right)
+4\alpha -\nu (t)$ one can neglect the rapidly oscillating terms in (\ref%
{dave}) and obtain the effective Hamiltonian \cite{palermo}%
\begin{eqnarray*}
\tilde{H} &=&E_{0}|R_{0}\rangle \langle R_{0}|+E_{2,-\mathcal{D}}|R_{2,-%
\mathcal{D}}\rangle \langle R_{2,-\mathcal{D}}| \\
&&+(\lambda e^{-it\left( E_{2,-\mathcal{D}}-E_{0}-\nu (t)\right) }|R_{2,-%
\mathcal{D}}\rangle \langle R_{0}|+h.c.)
\end{eqnarray*}%
\begin{equation*}
\lambda =\frac{i\mathcal{D}g }{2\Delta _{+}}\varepsilon _{\Omega }^{(k)}\exp ( -i\phi
^{(k)})~.
\end{equation*}%
Performing the time-dependent unitary transformation%
\begin{eqnarray}
U_{1}&=&\exp \left\{ -i\left[ \left( E_{0}+\frac{\nu (t)}{2}\right)
t|R_{0}\rangle \langle R_{0}|\right.\right. \nonumber \\
&&\left.\left.+\left( E_{2,-\mathcal{D}}-\frac{\nu (t)}{2}%
\right) t|R_{2,-\mathcal{D}}\rangle \langle R_{2,-\mathcal{D}}|\right]
\right\} \nonumber
\end{eqnarray}%
we get the final effective Hamiltonian%
\begin{eqnarray}
H_{f} &=&\frac{\tilde{\nu}(t)}{2}\left( |R_{2,-\mathcal{D}}\rangle
\langle R_{2,-\mathcal{D}}|-|R_{0}\rangle \langle R_{0}|\right)  \nonumber\\
&&+\left( \lambda |R_{2,-\mathcal{D}}\rangle \langle R_{0}|+h.c.\right)
\end{eqnarray}%
\begin{equation*}
\tilde{\nu}(t)\equiv \nu (t)+\dot{\nu}(t)t\,.
\end{equation*}
This is the standard Hamiltonian for a two-level system with a time-varying
energy splitting and constant off-diagonal coupling. For $\nu (t)=0$ the dynamics consists of periodic oscillation between the states $|R_{0}\rangle $ and $|R_{2,-\mathcal{D}%
}\rangle $, which approximately corresponds to the transition $|g,0\rangle\leftrightarrow |e,1\rangle $. On the other hand, if $\nu (t)=|\xi |t$ and $t$ is varied from $%
-\infty $ to $+\infty $ one can use the well known formula for the
{\it Landau-Zener} transition \cite{landau,zener,majo,stuck}. If for $t= -\infty $ the initial state was
$|R_{0}\rangle $, then for $t\rightarrow +\infty $ the probability of
occupancy of the state $|R_{2,-\mathcal{D}}\rangle $ is

\begin{equation}
P=1-\exp \left( -\frac{\pi |\lambda |^{2}}{|\xi |}\right)\,. \label{Dau}
\end{equation}%
Therefore if $|\dot{\nu}|\lesssim |\lambda |^{2}$ one can accomplish an
almost complete transition from $|R_{0}\rangle $ to $|R_{2,-\mathcal{D}%
}\rangle $. As shown in \cite{palermo} similar conclusion holds even if $\nu (t)$
varies within a finite interval around $\nu (t)=0$.

\emph{DCE regime}. For the modulation frequency
$\eta ^{(k)}=2\omega _{0}+2\left( \delta _{-}-\delta _{+}\right) -4\alpha
-\nu (t)$ we obtain the effective DCE Hamiltonian%
\begin{eqnarray}
H_{f}&=&\sum_{m=0}^{n_{\max }}\left( \frac{\tilde{\nu}(t)}{2}m-\alpha m\left( m-2\right) \right) |R_{m,\mathcal{D}%
}\rangle \langle R_{m,\mathcal{D}}|\nonumber \\
&&+\sum_{m=0}^{n_{\max }}\left( \lambda
_{m}|R_{m+2,\mathcal{D}}\rangle \langle R_{m,\mathcal{D}}|+h.c.\right) \label{igo}
\end{eqnarray}%
\begin{equation*}
\lambda _{m}=-\frac{i\delta _{-}}{2{\Delta _{+}}}\sqrt{\left( m+1\right) \left(
m+2\right) }\varepsilon _{\Omega }^{(k)}\exp
( -i\phi ^{(k)})\,.
\end{equation*}%
where $n_{\max }$ is the maximum number of excitations allowed by the
dispersive approximation and we denote $|R_{0,\mathcal{D}}\rangle \equiv
|R_{0}\rangle .$ As shown in \cite{igor} the Kerr term $\alpha m\left( m-2\right)$ in (\ref{igo}) is responsible for non-periodic collapse-revival behavior of $\langle n \rangle$.

\emph{JC Regime}. For the modulation frequency $\eta _{J}^{(k)}=\left\vert \Delta _{-}-2\delta _{+}J\right\vert +2\left\vert
\delta _{-}\right\vert J-2\left\vert \alpha \right\vert J^{2}-\nu$, with $J\ge 1$, we obtain the effective Hamiltonian (see \cite{palermo} for the validity range)
\begin{eqnarray}
H_{f} &=&\mathcal{D}\frac{\tilde{\nu}(t)}{2}\left( |R_{J,\mathcal{D}}\rangle
\langle R_{J,\mathcal{D}}|-|R_{J,-\mathcal{D}}\rangle \langle R_{J,-\mathcal{%
D}}|\right)  \nonumber \\
&&+(\lambda _{J}|R_{J,\mathcal{D}}\rangle \langle R_{J,-\mathcal{D}}|+h.c.)
\end{eqnarray}%
\begin{equation}
\lambda _{J}=-\frac{ig_{0}}{2\Delta _{-}}\sqrt{J}\varepsilon _{\Omega
}^{(k)}\exp (-\mathcal{D}i\phi ^{(k)}).
\end{equation}
This corresponds roughly to the transition between the states $|g,J\rangle$ and $|e,J-1\rangle$, reliant only on the rotating terms in the Hamiltonian.

\emph{ADCE Regime}. For the modulation frequency $\eta
_{J}^{(k)}=3\omega -\Omega _{0}+2\left( \delta _{-}-\delta _{+}\right)
\left( J-1\right) -2\alpha \left( J^{2}-2J+2\right) -\nu (t)$, where $J\geq 3
$, we obtain the effective Hamiltonian%
\begin{eqnarray}
H_{f}&=&\frac{\tilde{\nu}(t) }{2}\left( |R_{J,%
\mathcal{D}}\rangle \langle R_{J,\mathcal{D}}|-|R_{J-2,-\mathcal{D}}\rangle
\langle R_{J-2,-\mathcal{D}}|\right) \nonumber \\
&&+\left( \lambda _{J}|R_{J,\mathcal{D}%
}\rangle \langle R_{J-2,-\mathcal{D}}|+h.c.\right)
\end{eqnarray}%
\begin{equation}
\lambda _{J}=-\frac{i\mathcal{D}g\delta _{-}}{2\Delta _{+}\Delta_-}\sqrt{J\left( J-1\right) (J-2)}\varepsilon _{\Omega }^{(k)}\exp
( -i\phi ^{(k)})\,. \label{cd}
\end{equation}%
A more rigorous calculation performed in \cite{igor} resulted in the expression
\begin{equation*}
\lambda _{J}^{\prime }=\lambda _{J}\left( \frac{2\omega -\Delta _{-}}{%
2\omega +\Delta _{-}}\frac{\omega +\Delta _{-}}{\omega }\right)
\end{equation*}%
instead of (\ref{cd}). For the parameters of this paper, $\Delta_-=0.4\omega$, we obtain $\lambda _{J}^{\prime }/\lambda
_{J}=14/15\approx 0.93$, so the difference between $\lambda _{J}$ and $%
\lambda _{J}^{\prime }$ is insignificant and becomes smaller as the ratio $%
|\Delta _{-}|/\omega $ decreases.

ADCE corresponds to the induced coupling between the dressed states $|R_{J,\mathcal{D%
}}\rangle $ and $|R_{J-2,-\mathcal{D}}\rangle $, or approximately $|g,J\rangle $ and $|e,J-3\rangle $. In
particular, for $\nu =0$ we can solve the unitary dynamics for any initial
state by writing the density matrix in the dressed basis%
\begin{equation*}
\rho (t)=\sum_{\mathcal{S},\mathcal{T}}\sum_{n,l=0}^{\infty }\rho _{n,l}^{%
\mathcal{S},\mathcal{T}}(t)|R_{n,\mathcal{S}}\rangle \langle R_{l,\mathcal{T}%
}|
\end{equation*}%
If initially the off-diagonal elements ${\rho}_{n-2,n}^{\mathcal{S},%
\mathcal{-S}}(0)$ are zero, as occurs for the thermal state considered in this work, the solution
for the affected dressed states reads%
\begin{eqnarray*}
\rho _{J,J}^{\mathcal{D},\mathcal{D}}&=&\rho _{J,J}^{\mathcal{D},%
\mathcal{D}}( 0) \cos ^{2}(\vert \lambda
_{J}\vert t) +\rho _{J-2,J-2}^{\mathcal{-D},\mathcal{-D}}(
0) \sin ^{2}(\vert \lambda _{J}\vert t)  \\
\rho _{J-2,J-2}^{\mathcal{-D},\mathcal{-D}} &=&\rho _{J-2,J-2}^{\mathcal{%
-D},\mathcal{-D}}( 0) \cos ^{2}( \vert \lambda
_{J}\vert t) +\rho _{J,J}^{\mathcal{D},\mathcal{D}}(
0) \sin ^{2}( \vert \lambda _{J}\vert t),
\end{eqnarray*}%
while other probabilities remain unaltered. Therefore, one is able to annihilate two system excitations provided $\rho
_{J,J}^{\mathcal{D},\mathcal{D}}\left( 0\right) >\rho _{J-2,J-2}^{\mathcal{-D%
},\mathcal{-D}}\left( 0\right) $, or approximately when the
initial population of $|g,J\rangle $ is bigger than $%
|e,J-3\rangle $.

%
\begin{acknowledgments}
A. V. D. acknowledges partial support from the Brazilian agency Conselho Nacional de Desenvolvimento Cient\'ifico e Tecnol\'ogico (CNPq).
D. V. acknowledges support from the Brazilian agency Coordena\c c\~ao de Aperfei\c coamento de Pessoal de N\'ivel Superior (CAPES) through grant No. 88881.120135/2016-01.
\end{acknowledgments}

%
%


\begin{thebibliography}{99}
\bibitem{foot} Work can also be extracted from the cavity thermal state using the JC resonance, but we verified that the maximal amount of work is $3$ times smaller that in the ADCE regime.
\end{thebibliography}
\end{document}